\documentclass[sigplan,nonacm]{acmart}
\renewcommand\footnotetextcopyrightpermission[1]{}

\usepackage[utf8]{inputenc}

\usepackage{xspace} 
\usepackage{cleveref} 
\usepackage{hyperref} 
\usepackage{soul} 
\usepackage{graphicx} 
\usepackage{enumitem}
\usepackage{array,multirow}

\usepackage[frozencache=true,cachedir=./minted]{minted} 

\graphicspath{ {./images/} }

\usepackage{caption} 
\usepackage{subcaption} 

\newcommand{\sys}{Netherite\xspace}

\newcommand{\str}{{\color{red}\textbf{\texttt{*}}}} 
\newcommand{\kk}[1]{[{\color{cyan}kk: #1}]}
\newcommand{\seb}[1]{[{\color{green}sb: #1}]}
\newcommand{\dajusto}[1]{[{\color{blue}dj: #1}]}
\newcommand{\csm}[1]{[{\color{purple}csm: #1}]}

\renewcommand{\str}{} 
\renewcommand{\kk}[1]{}
\renewcommand{\seb}[1]{}
\renewcommand{\dajusto}[1]{}
\renewcommand{\csm}[1]{}
\renewcommand{\hl}[1]{{\color{red}#1}}

\newcommand{\hide}[1]{}

\newcommand{\eestarted}{\textsl{ExecutionStarted}}
\newcommand{\eecompleted}{\textsl{ExecutionCompleted}}
\newcommand{\etscheduled}{\textsl{TaskScheduled}}
\newcommand{\etcompleted}{\textsl{TaskCompleted}}
\newcommand{\mi}[1]{\mintinline{javascript}{#1}}

\newcommand{\resultpar}[1]{\paragraph{#1}\ }

\title{Serverless Workflows with Durable Functions and Netherite}

\author{Sebastian Burckhardt}
\affiliation{Microsoft Research}
\email{sburckha@microsoft.com}
 
\author{Chris Gillum}
\affiliation{Microsoft Azure}
\email{cgillum@microsoft.com}

\author{David Justo}
\affiliation{Microsoft Azure}
\email{dajusto@microsoft.com}

\author{Konstantinos Kallas}
\affiliation{University of Pennsylvania}
\email{kallas@seas.upenn.edu}

\author{Connor McMahon}
\affiliation{Microsoft Azure}
\email{comcmaho@microsoft.com}

\author{Christopher S. Meiklejohn}
\affiliation{Carnegie Mellon University}
\email{cmeiklej@cs.cmu.edu}

\begin{document}

\begin{abstract}
Serverless is an increasingly popular choice for service architects because it can provide elasticity and load-based billing with minimal developer effort. A common and important use case is to compose serverless functions and cloud storage into reliable workflows. However, existing solutions for authoring workflows provide a rudimentary experience compared to writing standard code in a modern programming language. Furthermore, executing workflows reliably in an elastic serverless environment poses significant performance challenges.

To address these, we propose Durable Functions, a programming model for serverless workflows, and Netherite, a distributed execution engine to execute them efficiently. Workflows in Durable Functions are expressed as task-parallel code in a host language of choice. Internally, the workflows are translated to fine-grained stateful communicating processes, which are load-balanced over an elastic cluster. The main challenge is to minimize the cost of reliably persisting progress to storage while supporting elastic scale. Netherite solves this by introducing partitioning, recovery logs, asynchronous snapshots, and speculative communication. 
 
Our results show that Durable Functions simplifies the expression of complex workflows, and that Netherite achieves lower latency and higher throughput than the prevailing approaches for serverless workflows in Azure and AWS, by orders of magnitude in some cases.
\end{abstract}
\maketitle

\section{Introduction}

Cloud service developers today have a choice: they may prefer to control the provisioning, maintenance, and recovery of servers explicitly, or they may prefer a \emph{serverless} architecture where applications are layered on top of services that manage servers automatically. 

The term \emph{serverless} is often considered synonymous with \textit{Functions-as-a-Service} (FaaS), which was pioneered by Amazon~\cite{aws-lambda} and is now ubiquitous \cite{google-cloud-functions,azure-functions,openwhisk,fission}. In FaaS, a function is a piece of application code designed to respond to an individual event. Compared to a virtual machine or a compute instance, a function is significantly more fine-grained and can be scheduled for execution much faster on a pool of compute resources. Furthermore, FaaS platforms support per-invocation billing. This means that a service built on FaaS is not only highly available, but is both (1) very cheap to operate under low load, and yet (2) can scale automatically to a high load, at a proportional cost. Given the potential developer productivity boost that the serverless paradigm provides, it is anticipated to become increasingly prominent for cloud applications \cite{DBLP:journals/corr/abs-1902-03383,revolution}. 

Developing complex stateful serverless applications with FaaS is not straightforward, however, as it provides limited execution guarantees \cite{jangda-et-al}. For example, a function may be restarted many times before completing, and must complete within strict time limits. 

Workflows have been identified by many~\cite{sreekanti2020cloudburst,eismann2020predicting,azure-durable-functions,step-functions,scientific-workflows} to be the missing link to enable the development of full-featured serverless applications.
What differentiates workflows from simple function composition is that they provide stronger execution guarantees. Unfortunately, developing workflows for a serverless environment still poses significant performance and programmability challenges. We now discuss these challenges and how we address them.


\paragraph{Workflows as Code.}
Workflows have been used for several decades, in various shapes and forms.
One approach is \emph{unstructured composition}, where all control flow is explicit. For example, \emph{triggers} are a common composition primitive for FaaS: to sequence two functions, the first function can write to a file or queue, which then triggers the execution of a second function. Two major drawbacks of unstructured composition are (i) that it doesn't support all forms of composition, such as value aggregation, and (ii) that the control flow is dispersed over many places, reminiscent of "spaghetti code". Another approach is \emph{structured composition}, where the system provides higher-level control flow abstractions, such as sequencing, sequential and parallel iteration, and error handling. Structured composition is often achieved through restricted declarative schemas, such as XML~\cite{shegalov2001xml}, JSON~\cite{step-functions}, or visual design tools~\cite{azure-logic-apps}.

In contrast, Durable Functions (DF), our proposed programming model, achieves structured composition expressed \emph{as code}, in a standard programming language of choice (such as  JavaScript, Python, C\#, or PowerShell). The benefit over declarative approaches is that DF workflows can take advantage of all the familiar control flow abstractions and the ecosystem of libraries and tools of a mature host language. 
DF persists the intermediate state of a workflow using record and replay.
\hide{
Therefore, it does require developers to keep the orchestration code deterministic (to avoid divergence during replay) and of bounded length.
}

\paragraph{Serverless Computation Model.}

In order to keep the engine development separate from the programming model we propose a computation model that contains two simple "serverless primitives": \emph{stateless tasks} and \emph{stateful instances}. This acts as an interface between the programming model and the execution engine: DF is translated into the computation model by encoding workflows as stateful instances, and \sys implements it. This separation allows independent experimentation on the programming or the engine part---in fact, we benefited from this separation since \sys was built as a replacement for the existing Durable Functions implementation. The model is also designed to facilitate elasticity: tasks and instances are both fine-grained and communicate via messages, which makes it possible to dynamically load-balance them over an elastic cluster.


\paragraph{Causally Consistent Commit.}

A common challenge for workflow systems is to articulate a reliability guarantee that is strong, easy to understand for programmers, and efficiently implementable. To this end, we define a guarantee called \emph{causally consistent commit} (CCC) using execution graphs. It is stronger than "at-least-once" or "effectively-once", and more realistic than "exactly-once". In essence, it guarantees atomicity: a step that fails is \emph{aborted}, along with all steps that causally depend on it. 
\hide
{
\kk{Maybe replace more concrete with more realistic? I guess that we want to argue that it is more efficiently implementable (in contrast to exactly once which is not implementable at all in general).} 
\seb{agreed.}
\kk{I think that calling them not sufficiently precise or comprehensive is not totally true. There must be more precise definitions of such terms. Instead I think that we need to focus on the fact that effectively once lies somewhere in between exactly once and at-least-once, being both strong, but also achievable in practice without huge overheads. Our contribution is essentially identifying this gap between the two, formulating a guarantee that strikes a nice balance.}
}

\paragraph{Batch Commit.}

In order to guarantee reliability, workflow solutions need to persist workflow steps in storage. This is commonly achieved by persisting the state and steps of each workflow individually\footnote{This is the case with unstructured composition, as well as the existing DF implementation.}, creating a throughput bottleneck due to the limited number of I/O operations storage can handle per second.
To avoid this problem, we designed \emph{\sys} so it can persist many steps, by different workflow instances, using a single storage update. This is achieved by grouping the fine-grained instances and tasks into partitions. Each partition can then persist a batch of steps efficiently by appending it to its commit log in cloud SSD storage.

\paragraph{Speculation Optimizations. } 

A conservative workflow execution engine would wait until a step is persisted before proceeding with the next step. This introduces a significant latency overhead since storage accesses are on the critical execution path. We show that with careful \emph{local} and \emph{global} speculation, \sys moves these storage accesses off the critical path, significantly reducing latency, while still providing the CCC guarantee.\str

\paragraph{Elastic Partition Balancing.} 

\sys uses a fixed number of partitions (32) that communicate via a reliable ordered queue service. It can move individual partitions between nodes, by persisting and then recovering their state on a different node. In particular, it can re-balance the partitions as needed. For example, on a one-node cluster, all 32 partitions are loaded on a single node. On a four-node cluster, each node has eight partitions, and so on, up to 32 nodes with one partition each. \sys can also scale to zero if the application is idle: on a zero-node cluster, all partitions reside in cloud storage.

\paragraph{Evaluation.} 

Our evaluation on five workflows, two of which are taken from real applications, indicate that the DF programming model offers significant benefits regarding development effort. 
In particular, the availability of general loops, exception handling, and functional abstraction (provided by the host language) greatly improve the experience when dealing with complex workflows. 

Yet, the benefits are not limited to the developer experience: the execution performance with Netherite is better than with common serverless alternatives, \emph{across the board}. For instance, Netherite orchestrations outperform  trigger-based composition by orders of magnitude, both on AWS and Azure. They also exhibit better throughput and latency than the current Durable Functions production implementation, by an order of magnitude in some situations. Finally, a workflow composing AWS lambdas completes faster in \sys (deployed in Azure and invoking lambdas through HTTP) rather than in Step Functions (deployed in AWS and invoking lambdas directly). 

\subsection{Contributions}
 
We make the following contributions:

\begin{itemize}
\item We introduce the Durable Functions Programming Model, which allows code-based structured expression of workflows in multiple languages (\S\ref{sec:durablefunctions}).
\item We demonstrate how to break down complex workflows into just two serverless primitives, and define the causally-consistent-commit guarantee (\S\ref{sec:model}).
\item We provide an architecture and implementation that realize these concepts (\S\ref{sec:netherite}) and demonstrates the power of speculation optimizations (\S\ref{sec:optimizations}).
\item We evaluate the Durable Functions programming model and \sys implementation on several benchmarks and case studies, comparing it to commonly used serverless composition techniques (\S\ref{sec:evaluation}).
\end{itemize}
%
%
Overall, our contributions bring the development of complex full-fledged serverless applications within reach: providing cloud developers with (i) Durable Functions, a mature programming environment that allows them to have their application in one place; and (ii) \sys, an efficient execution engine that provides strong reliability guarantees.
\section{Durable Functions}\label{sec:durablefunctions}

DF is a programming model that offers a novel combination of abstractions for reliable workflows. It supports both simple scenarios, such as workflows of tasks that perform sequential or parallel composition and iteration, as well as advanced concepts, such as durable entities and critical sections. Its implementation is open-source \cite{df-extension-repo,df-js-repo,df-python-repo, df-powershell-repo}, and is built on top of the Azure Functions framework \cite{azure-functions-repo}. The currently supported languages are JavaScript, Python, C\#, and PowerShell.  

\vspace{.07in}\noindent\textbf{Orchestrations} 
are reliable workflows written in a task-parallel style. An example illustrating a simple orchestration, a sequential composition of two functions, is shown in Fig.~\ref{fig:simplesequence}. Lines 1--3 declare that this is an orchestration function named \texttt{SimpleSequence}. When invoked, this orchestration reads its input (line 7) and then calls an activity with name \texttt{F1}. The term "activity" is DF terminology for a stateless serverless function, that can take an input and return an output. We have omitted the code for the activities in our examples. The \texttt{await} on line 8 indicates that the orchestration should resume execution only after \texttt{F1} is complete. The returned result is then passed to the next function \texttt{F2} (line 9). When the latter finishes, the orchestration returns the final result (line 10). If anything goes wrong, the exception handler (line 13) can take appropriate action.

A slightly more interesting example containing parallel iteration is shown in Fig.~\ref{fig:orchestration}. It shows a JavaScript example of an orchestration that creates thumbnails for all pictures in a directory. It receives a directory name as input (line 4), and then calls an activity "GetImageList" (line 6) to obtain the list of files. The \texttt{yield} on line 6 serves as a JavaScript equivalent of \texttt{await}. Next, to create the thumbnails in parallel, the orchestration starts an activity for each of them, without \text{yield}, thus not waiting for the result, but storing the tasks in an array (line 12). Next it calls \texttt{yield} to indicate that the orchestration should resume after all the parallel tasks are complete (line 16). Finally, it aggregates (sums) all the returned numbers (sizes) and returns the result (line 18).
    
\begin{figure}
\begin{minted}[fontsize=\footnotesize,linenos]{csharp}
[FunctionName("SimpleSequence")]
public static async Task<int> Run(
    [OrchestrationTrigger] IDurableOrchestrationContext c)
{
    try
    {
        var x = c.GetInput<int>();
        var y = await c.CallActivityAsync<int>("F1", x);
        var z = await c.CallActivityAsync<int>("F2", y);
        return z;
    }
    catch (Exception) {
        // Error handling or compensation can go here.
    }
}
\end{minted}
    \caption{Sequencing two functions F1 and F2 using a durable functions orchestration in C\#.}
    \label{fig:simplesequence}
\end{figure}
    
\begin{figure}
\begin{minted}[fontsize=\footnotesize,linenos]{javascript}
const df = require("durable-functions");
module.exports = df.orchestrator(function*(context) {
  // Get the directory input argument
  const directory = context.df.getInput();
  // Call an activity and wait for the result
  const files = yield context.df.callActivity(
    "GetImageList", directory);
  // For each image, call activity without waiting
  // and store the task in a list
  const tasks = [];
  for (const file of files) {
    tasks.push(context.df.callActivity(
      "CreateThumbnail", file));
  }
  // wait for all the tasks to complete
  const results = yield context.df.Task.all(tasks);
  // return sum of all sizes
  return results.reduce((prev, curr) => prev + curr, 0);
});
\end{minted}
\caption{Example orchestration using the Durable Functions JavaScript API. It calls
an activity \texttt{GetImageList}, and then, in parallel, \texttt{CreateThumbnail} for each image. It then waits for all to complete and returns the aggregated size.}
\label{fig:orchestration}
\end{figure}

\begin{figure}
\begin{minted}[fontsize=\footnotesize,linenos]{csharp}
public class Account
{
    public int Balance { get; set; }
    public int Get() => Balance;
    public void Modify(int Amount) { Balance += Amount; }

    // boilerplate for Azure Functions (feel free to ignore)
    [FunctionName(nameof(Account))]
    public static Task Run([EntityTrigger] 
      IDurableEntityContext ctx)
        => ctx.DispatchAsync<Account>();
}
\end{minted}
\caption{Example entity using the Durable Functions C\# API. Its state 
is an integer \texttt{Balance}, and
it has operations \texttt{Get} and \texttt{Modify} to read or update it.}
\label{fig:entity}
\end{figure}

\vspace{.07in}\noindent\textbf{Entities} 
are addressable units that can receive operation requests and execute them sequentially and reliably.
\hide{
\kk{I am not sure whether persisting state should be mentioned here. Reliability (which is a guarantee that DF provides) should be mentioned though.}
}
Fig.~\ref{fig:entity} shows a C\# example of an entity representing a bank account. The state of the entity is an integer (line 3) that is read by an operation \texttt{Get} (line 4) and updated by an operation \texttt{Modify} (line 5). An entity ID consists of two strings, the entity name and the entity key. For example, an account entity may be identified by ("Account", "0123-44918"). All entity operations are serialized, that is, their execution does not overlap, which provides a simple solution to basic synchronization challenges. The concept of entities is similar to virtual actors, or \emph{grains}, as introduced by Orleans~\cite{bykov2011orleans}.  



\vspace{.07in}\noindent\textbf{Critical sections} help to address synchronization challenges involving durable state stored in more than one place, such as in multiple entities and/or in external services. For example, consider an orchestration that intends to transfer money between accounts. Fig.~\ref{fig:transfer} shows such an orchestration, using the C\# API. First, we obtain the input parameters (source, destination, and amount) on line 5. Then, we construct entity IDs for the two accounts (line 7, 8). The \texttt{LockAsync} call on line 10 locks both account entities for the duration of the critical section (lines 11 through 23), enforcing exclusive access. On line 12, we read the current balance of the source account by calling the \texttt{Get} operation.\footnote{Our .NET interface also offers an interface-and-proxy-based syntax for calling  entities that provides type checking for operations and arguments. It takes a bit more space so we chose the untyped syntax for this small demonstration example.} If the balance does not cover the amount (line 13) we return false (line 15), otherwise we modify both accounts by calling the two account entities in parallel (lines 20, 21). After both entities finish the operation, the \texttt{await} (line 19) completes and we return true, exiting the critial section, and releasing both locks.

\begin{figure}
\vspace{.1in}
\begin{minted}[fontsize=\footnotesize,linenos]{csharp}
[FunctionName("Transfer")]
public static async Task<bool> Transfer(
    [OrchestrationTrigger] IDurableOrchestrationContext ctx)
{
  (string source, string dest, int amount) = 
    ctx.GetInput<string,string,int>();
  EntityId sourceId = new EntityId("Account", source);
  EntityId destId = new EntityId("Account", dest);
  
  using (await ctx.LockAsync(sourceId, destId))
  {
    int bal = await ctx.CallEntityAsync<int>(sourceId, "Get");
    if (bal < amount)
    {
      return false;
    }
    else
    {
      await Task.WhenAll(
        ctx.CallEntityAsync(sourceId, "Modify", -amount),
        ctx.CallEntityAsync(destId, "Modify", +amount));
      return true;
} } }
\end{minted}
\vspace{-.17in}\caption{Example of an orchestration with a critical section that reliably transfers money between account entities.}
\label{fig:transfer}
\end{figure}

\begin{figure}
\begin{center}
\begin{math}\small
\noindent\begin{array}{|c|c|c|@{\quad}l}
    \rotatebox[origin=c]{90}{\text{\ step 1}} 
  & \rotatebox[origin=c]{90}{\text{\ step 2}} 
  & \rotatebox[origin=c]{90}{\text{\ step 3}}  \\\hline
x & x & x &\eestarted(\mi{"SimpleSequence"}, \mi{x})\\
x & x & x &\etscheduled(\mi{"F1"}, \mi{x})\\
  & x & x &\etcompleted(\mi{y})\\
  & x & x &\etscheduled(\mi{"F2"}, \mi{y})\\
  &   & x &\etcompleted(\mi{z})\\
  &   & x &\eecompleted(\mi{z})\\
\end{array}
\end{math}
\end{center}
\caption{The history recorded for the orchestration in Fig.~\ref{fig:simplesequence}.}\label{fig:history}
\end{figure}

\subsection{Orchestration Persistence}
In contrast to stateless functions, orchestrations do not have to remain in memory, accumulating billing charges, while they wait for a step to complete. Instead their progress can be stored in durable storage and retrieved when the step has completed. This is particularly important for long-running workflows.

Rather than persisting the program location, variables, and heap, DF records a \emph{history} of events. For example, the orchestration from Fig.~\ref{fig:simplesequence} executes in three steps, with partial histories as shown in Fig.~\ref{fig:history}. It is possible to re-hydrate the intermediate state of an orchestration from storage by replaying the persisted partial history. Completed tasks are not re-executed during replay; rather, the recorded results are reused.

Replay can cause problems if the orchestration contains nondeterminism or if histories are excessively long. Developers are expected to avoid these issues by (1) encapsulating  nondeterminism in activities, and (2) using sub-orchestrations, or restarting orchestrations, to limit history size. DF also includes a static analysis tool that can detect common mistakes of this kind for its C\# front end.

\subsection{Comparing DF to Triggers}

Perhaps the simplest way to construct a workflow is to specify \emph{triggers}, also called \emph{bindings}, which launch functions in response to storage events. For example, to implement a simple sequence, we can instruct each step to write its result to storage (typically, a queue or a file) which then triggers the next step. Because of their conceptual simplicity and wide availability, triggers are commonly used by FaaS developers. 

Authoring complex workflows using just triggers is possible, but the developer experience is not ideal. One has to create bindings and queues/directories for each intermediate step of the workflow, which is tedious and error-prone. Also, triggers do not support specifying that a step should wait for the completion of multiple previous steps, which is a common requirement (e.g. on line 16 in Fig.~\ref{fig:orchestration}). Finally, triggers do not offer a convenient way to specify error handling (e.g. as in line 13 in Fig.~\ref{fig:simplesequence}).

\subsection{Comparing DF to Step Functions}
%
%
Another alternative for specifying workflows declaratively is to define a state machine that invokes and guides the composition of functions. For example, with AWS Step Functions \cite{aws-step-functions}, a computation graph is expressed using a JSON-based declarative DSL. Nodes in this graph either invoke a serverless function (or some other cloud service), forwarding its output to a specified target node, or they inspect their input to make a conditional state transition. Special nodes exist to handle exceptions or invoke other AWS services.

Conceptually, an AWS Step Function is comparable to a DF orchestrator. But the programming model is different: DF expresses orchestrations in a mainstream language, while Step Functions use a JSON-based schema with limited options for abstraction and control flow. For example, sequential loops with dependent iterations cannot be expressed (\S\ref{ssec:eval-programmability}).

\section{Computation Model}
\label{sec:model}

In this section we describe the core of the serverless computation model that underlies Durable Functions and is implemented by Netherite. By describing this model abstractly using execution graphs, we provide a solid foundation that allows us to state and explain the "causally-consistent-commit" execution guarantee of Netherite. 




\subsection{Tasks and Instances}\label{sec:tasksandinstances}

Computations in our model are built from tasks and instances that communicate via messages. We distinguish two types of messages:
\begin{itemize}
    \item \emph{Task messages} are used to start a task. Tasks are stateless and can be executed anywhere. When a task finishes executing, it sends a single result message.
    \item \emph{Instance messages} target a specific stateful instance, identified by an \emph{instance ID}. When processed, an instance message may read and update the state of that instance, and may also produce additional messages.
\end{itemize}

\hide{
\kk{I am not exactly sure why this is mentioned here. It seems to be a bit detached from the rest of the narrative. Maybe we can say that the model helps in addressing a fundamental challenge for workflow systems (that their steps need to be atomic).}
\seb{I have removed it since it is explained in other places also.}
}
%
%
The fine granularity of tasks and instances, and the state encapsulation afforded by the message passing paradigm, facilitate elasticity as they allow us to balance task and instance execution across an elastic cluster. For stateless tasks, load balancing is straightforward. For stateful instances, it requires a bit more work. We describe our solution in \S\ref{sec:netherite}. 


\subsection{Execution graphs}

\begin{figure}
    \centering
    \includegraphics[width=.8\columnwidth]{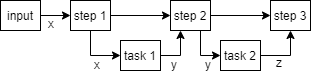}
    \caption{Execution graph for a simple sequence of two tasks as in Fig.~\ref{fig:simplesequence}. Vertices are labeled to indicate the vertex type, and message edges are labeled with the value propagated.}
    \label{fig:ss-graph}
\end{figure}

To visualize execution states and execution histories, we use \emph{execution graphs}. There are three types of vertices:
\begin{itemize}
\item An \emph{input} vertex represents an external input message. 
\item A \emph{task} vertex represents a stateless task.
\item A \emph{step} vertex represents the processing of a batch of one or more messages by a stateful instance.
\end{itemize}
We call the task and step vertices \emph{work items}, since both represent the processing of messages. Edges in the graph represent direct causal dependencies:
\begin{itemize}
    \item A \emph{message} edge from $v_1$ to $v_2$ means that $v_2$ consumes a message produced by $v_1$. 
    \item A \emph{successor} edge from $v_1$ to $v_2$ means that they are successive steps of the same instance.
\end{itemize}
For an example, see Fig.~\ref{fig:ss-graph}. This execution graph corresponds to the simple sequence from Fig.~\ref{fig:simplesequence}. The input is the message that starts the orchestration; the orchestration then proceeds in three steps: (1) receive input and issue first task, (2) receive first task result and issue second task, and (3) receive second task result and finish.

We call an execution graph \emph{consistent} if it is consistent with a sequential execution of atomic processing steps as described in \S\ref{sec:tasksandinstances}. We call an execution graph \emph{complete} if all messages produced are also consumed.

\subsection{Faults and Recovery}\label{sec:faults}


A critical reality of service-oriented environments is the prevalence of faults: tasks may time out, nodes may crash (e.g., run out of memory) and reboot, and service connections may be temporarily interrupted. For example, attempts to persist instances to storage, to send a message, or to acknowledge the receipt of a queue message, may fail intermittently. 

What does it mean for a workflow execution to be correct in the presence of faults and recovery? Ideally, faults would be invisible. This is sometimes called an "exactly-once" guarantee, since it means that each message is processed exactly once. In general it is unfortunately not possible to implement this guarantee.
\hide{
\kk{I think that this needs to be qualified since it is theoretically possible if one has control over all services and can implement transactions over different services. We could say "...not possible to implement this guarantee in general, for any combination of external services".}
\seb{I think the statement is already sufficiently qualified with "in general". Also, this is not so much an argument about the consistency model in the sense that I don't say "observationally exactly once", but just that code actually must internally execute more than once. External services are not the only way to observe re-executions. System traces also. And resource consumption.}
}
The reason is that, when recovering from a crash, some progress may be lost, and some code must therefore be re-executed, possibly re-performing an irrevocable effect on an external service.

Because of that, many workflow systems settle for an "at-least-once" guarantee, where a message may be processed more than once, and thus its effects may also be duplicated. To handle duplicates correctly, developers usually employ a technique called "effectively-once" \cite{effectively-once-pulsar,effectively-once-tweet}: it combines the at-least-once guarantee with additional  mechanisms that ensure that all effects of processing a message are \emph{idempotent}. It may at first appear that the combination of at-least-once and idempotence is sufficient to hide faults. However, that is not true in the presence of nondeterminism. The reason is that if re-processing a message produces \emph{different effects} (e.g. sends a message to a different queue, or updates a different storage location), the effects of both executions remain, instead of being deduplicated.
\hide{
\csm{effectively-once already exists as an industry thing and has for a while -- can we actually propose it here?}\seb{We should perhaps use a different term; it seems that the existing use of this term is based on message deduplication which is not equivalent to our definition, if message processing is not deterministic. Maybe just "atomic" processing?}
\csm{see; https://twitter.com/viktorklang/status/789036133434978304 "I'm coining the phrase "effectively-once" for message processing with at-least-once + idempotent operations."}
\csm{I do like ``atomic'' processing better, but I'm not sure.}\seb{My current suggestion is "causally consistent commit", or CCC.}
}


\paragraph{Causally consistent commit.}
To address the shortcomings of the aforementioned guarantees we propose a guarantee called "causally-consistent-commit". 
The intuition behind it is that if we re-execute a work item, we have to ensure that all internal effects that causally depend on the previous execution are \emph{aborted}; in particular, any produced messages are discarded and updated instance states are rolled back. 

Note that work items are aborted only due to transient error conditions in the underlying execution infrastructure, never because of exceptions in the user code. A work item throwing a user-code exception is considered completed, with the exception being the result. 

As far as the system-internal state is concerned, CCC is observationally equivalent to "exactly-once". However, unlike "exactly-once", and like TM opacity \cite{opacity}, CCC gives semantics to aborted work items as well; it does not "pretend that they never happened". This is important because in reality, aborted work items cannot be completely hidden, but remain visible to users. For example, they may have external effects that cannot be undone (e.g. calls to external services), and they may appear in system traces and when debugging. 
\hide{
\kk{Can we bubble this paragraph higher up, maybe where we first describe work items?}
\seb{Done.}
}

\subsection{Fault-augmented execution graphs}

\begin{figure}[t]
    \centering
    \includegraphics[width=.6\columnwidth]{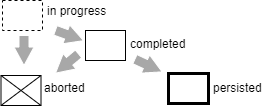}
    \caption{Progress states for work items.}
    \label{fig:progressstates}
\end{figure}

We now show how to augment execution graphs with a notion of progress, to precisely describe how to understand the effect of faults and recovery. We use the following \emph{progress states} (Fig.~\ref{fig:progressstates}) to mark the status of work items (= step or task vertices) in the execution graph.

\begin{description}
\item[In progress] This state represents a work item that is being executed. It is the initial state of a work item, and can change to \textbf{Completed} or \textbf{Aborted}.
\item[Completed] This state represents a work item whose execution has completed, but whose effects are not yet permanent. It can change to \textbf{Persisted} or \textbf{Aborted}.
\item[Persisted] This state represents a work item whose execution has completed and whose effects have been permanently persisted.
\item[Aborted] This state represents a work item that was permanently aborted. 
\end{description}

\begin{figure}
    \centering
    \begin{subfigure}[b]{\columnwidth}
        \centering
        \includegraphics[width=.75\columnwidth]{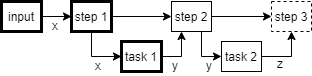}
        \caption{Execution graph immediately before the node crash.}
        \label{fig:pre-failure-execution}
    \end{subfigure}
    \begin{subfigure}[b]{\columnwidth}
        \centering
        \vspace{.2in}
        \includegraphics[width=.75\columnwidth]{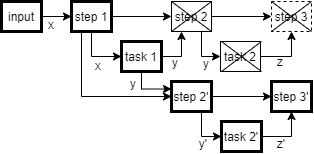}
        \caption{After the crash, recovery, and re-executing to completion.}
        \label{fig:post-failure-execution}
    \end{subfigure}
    \caption{Execution graphs for the simple sequence (Fig.~\ref{fig:simplesequence}) before and after a node crash.
    \hide{
    \kk{It might help if we move the aborted tasks below (essentially mirroring the way the (a) part of the figure looks).}
    \seb{Yes, I can fix this.}
    }
    }
    \label{fig:failure-execution}
\end{figure}

For an example, see Fig.~\ref{fig:failure-execution}, which illustrates what may happen to an incomplete execution of the simple sequence (Fig.~\ref{fig:simplesequence}) upon a crash. In \Cref{fig:pre-failure-execution} we show a graph representing the execution right before the crash. It has completed (but not yet persisted) step 2 and task 2, and is in the middle of executing step 3. In \Cref{fig:post-failure-execution} we show the final execution graph after the system recovers from the crash. Upon crash and recovery, step 2, task 2, and step 3 are aborted and then re-executed. Note that we do not assume determinism of tasks or steps: the re-executed task 2' may produce a different result $z'$ than the result $z$ returned by task 2 before the crash. This illustrates the importance of maintaining causal consistency through crashes and recovery. We now define this notion precisely.

\subsection{Causally-Consistent-Commit Guarantee}\label{sec:consistencyguarantees}

We say an implementation guarantees causally-consistent-commit (CCC) if it maintains consistency at each progress level. Concretely, let $\mathcal{I}$, $\mathcal{C}$, $\mathcal{P}$, and $\mathcal{A}$ be the subset of vertices in the respective progress states. Then, at all times, the subgraph $\mathcal{P}$, the subgraph $\mathcal{P}\cup\mathcal{C}$, and the subgraph $\mathcal{P}\cup\mathcal{C}\cup\mathcal{I}$ must each be a consistent execution graph.

In a CCC execution, the following always hold:
\begin{itemize}
\item a persisted work item causally depends only on work items that are also persisted.
\item a work item that causally depends on an aborted work item is also aborted.
\item each message is consumed by at most one non-aborted work item.
\item in a complete execution, each message is consumed by exactly one non-aborted work item.
\end{itemize}


\subsection{Speculation}\label{sec:defspeculation}

A benefit of the CCC guarantee is that it enables \emph{speculation}: a work item can proceed even if it causally depends on a not-yet-persisted work item. For example, in \Cref{fig:pre-failure-execution}, task 2 and step 3 are speculative. 

Speculation can boost performance because it allows the aggregation of storage accesses. Consider again the running example from Fig.~\ref{fig:ss-graph}. In a conservative implementation without speculation, we must save progress to storage after each work item. With speculation, however, we may execute the entire orchestration first and then persist all work items at once. Latency and throughput are both improved: a sequence of 5 storage accesses takes much longer, and consumes more system resources, than a single batched storage access.

\subsection{Correspondence with DF}
\label{sec:correspondence}

Our computation model can directly express programs in DF.
Stateful instances are used to represent DF orchestrations and entities, and tasks are used to represent DF activities. 

\section{\sys}
\label{sec:netherite}

In this section we introduce \sys, an execution engine that efficiently implements the computation model from \S\ref{sec:model}. We first give an overview of the \sys architecture and then explain how partitions are persisted in more detail.

The computation model from \S\ref{sec:model} uses fine-grained instances and tasks. A reliable tracking of messages and states for a large number of such instances creates significant overhead, however. To address this issue, \sys maps instances to partitions based on their identifier and then uses partitions as the unit of distribution and recovery. Using coarse-grained partitions, as opposed to fine-grained instances, also mitigates I/O bottlenecks by aggregating communication and storage accesses. 

\begin{figure}
    \centering
    \includegraphics[width=\columnwidth]{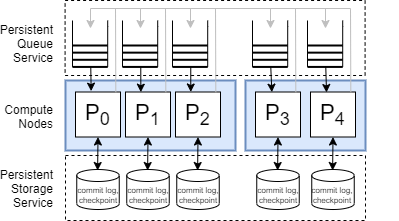}
    \caption{Illustration of the \sys architecture.}
    \label{fig:partitions}
\end{figure}
For an overview of the architecture, see Fig.~\ref{fig:partitions}. Partitions communicate via a persistent queue service; each partition has its own, ordered queue. The state of each partition is continuously saved to persistent storage, using an incremental commit log and occasional checkpoints. Storage leases are used to ensure that a partition is loaded on at most one compute node at a time.
This architecture addresses the following challenges:

\begin{description}

\item{\textbf{Partition recovery.}}  All messages are stored in a persistent queue; and all partition states are stored in persistent storage. If a partition crashes, we can recover it on a different compute node and resume processing at the correct input queue position (which is stored as part of the partition state).

\item{\textbf{Partition mobility.}}  Similarly, we can move a partition between compute nodes by shutting it down and then recovering it on a different node. 

\item{\textbf{Elasticity}} is a critical requirement in the serverless setting: we must support the addition and removal of compute nodes. This is achieved by using a sufficiently large number of partitions (32 by default), and re-balancing the partitions across compute nodes as needed. 

\item{\textbf{Batch commit.}} As explained in the introduction, the large number of storage writes can easily become a throughput bottleneck of a workflow processing system. We solve this problem by using a commit log, which makes it possible for partitions to persist state changes using a batch-append to the commit log. This reduces the number of I/O operations and can support high throughput, especially if backed by SSD storage.


\end{description}

\begin{figure}
    \centering
    \includegraphics[width=.65\columnwidth, viewport=2 17 196 240]{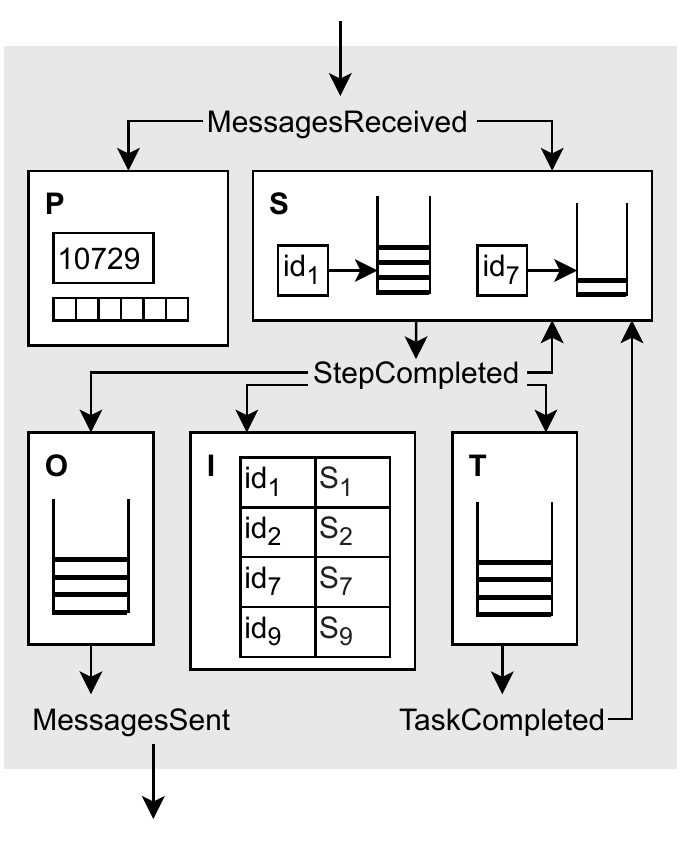}
    \caption{Illustration of the partition-internal state.\hide{\dajusto{I found this Figure and section to be quite confusing. I'm hoping we can either walk through an example in the text or work to clarify the technical description.}\seb{I have updated this section with an overview of the architecture and a figure, hope that helps.}}}
    \label{fig:partitionstate}
\end{figure}

\subsection{Partition State Persistence}\label{sec:espstate}

Efficiently saving the partition state to storage is a critical requirement for \sys. 
\hide{In general, there are many options to achieve this. We now describe the design choices.
}
To this end, \sys employs \emph{event sourcing}, a dual persistence model using a flexible combination of a commit log and checkpoints. With event sourcing, the partition state is a deterministic function of the sequence of events stored in the commit log. Changes to the partition state can thus be efficiently persisted by appending batches of events to the commit log, and the partition state can be recovered by replaying the events in the log. Additionally, partitions periodically take a checkpoint of their state, to reduce the number of events that have to be replayed on recovery.

\paragraph{State components. }Each partition must keep track of the state of all its instances, must send and receive messages, and must execute tasks and steps. To this end, the state of a partition includes these components (Fig.~\ref{fig:partitionstate}):
\begin{itemize}
    \item[I.] A map from instance IDs to instance states.  
    \item[P.] The queue position of the last processed input, and a deduplication vector.
    \item[S.] Buffers for incoming messages, by instance ID.
    \item[O.] A buffer for outgoing messages.
    \item[T.] A list of pending tasks.
\end{itemize}
Introducing buffers decouples the work for sending and receiving of messages, processing steps, and processing tasks, which in turn increases pipeline parallelism and enables batching. As required by the event sourcing paradigm, execution progress is recorded as a sequence of atomic events that update the partition state deterministically. There are 4 event types:
\hide{\csm{Minor readability point: this list might be improved if the message types were italicized or something.}\seb{ok}}
\begin{itemize}
    \item \textsl{MessagesReceived}. Updates P (advances position and deduplication vector) and S (enqueues messages).
    \item \textsl{MessagesSent}. This updates O (removes messages).
    \item \textsl{TaskCompleted}. This updates S (enqueues response) and T (removes completed task).
    \item \textsl{StepCompleted}. This updates I (updates instance state), S (removes consumed messages), O (adds produced messages), and T (adds produced tasks).
\end{itemize}

\paragraph{Instance State Caching.}
Keeping the state of all instances of a partition in memory is expensive and not always possible. Also, loading that state into memory on partition recovery is slow. Thus, it is important to have a caching  mechanism that keeps only the most recently used instances in memory, while the rest remains in storage. \sys achieves that by leveraging FASTER~\cite{chandramouli2018faster}, a hybrid key-value store that coexists in memory and storage. FASTER exploits temporal access patterns to keep "hot" keys in memory while evicting the rest in storage. It is implemented on top of a hybrid log, which allows it to perform fewer batched storage accesses.
\section{Optimizations}\label{sec:optimizations}

The baseline \sys implementation is conservative: the messages produced by a work item execution are first persisted to storage before being propagated. As explained in \S\ref{sec:defspeculation}, speculation can improve performance by moving this storage access off the critical path. This does not compromise the CCC guarantee, because we take care to properly propagate aborts along causal dependencies. We now describe two levels of speculation that are supported as optional optimizations in \sys.
\hide{
\kk{I wouldn't call them optional optimizations. But rather *the* optimized \sys implementation. There are two reasons for this: (1) the optimizations only give benefits in our evaluation, (2)I have had experience in the past where calling an optimization optional made reviewers ask for support by the system to choose which optimizations to enable and which not too.}
}

\paragraph{Local Speculation.} 

With local speculation, we allow messages to be processed immediately (before the work item is persisted) as long as the message stays within the same partition. 
Messages headed for different partitions are held up in the outbox $O$ until after their work item is persisted.
\hide{
\kk{I feel that we are not adequately motivating the benefits of speculation and the issue with the conservative solution. The explanation of the speculation therefore feels a little too dry and not that exciting, simply explaining how it works. I am not exactly sure how to address it though.}
}

Thus, we never need to propagate aborts to other partitions. Locally, within a single partition, aborts "automatically" respect causality because we use a single, causally consistent commit log to persist the partition state. After a crash, the partition state reverts to the persisted prefix of the commit log, which implicitly aborts all non-persisted work items.

Local speculation provides significant benefits for independent workflows that do not communicate with other stateful instances, therefore staying within a single partition during their execution. 
That includes the common case of orchestration workflows that only use a single instance and compose multiple tasks, such as the examples in Fig.~\ref{fig:simplesequence} and Fig.~\ref{fig:orchestration}. 


\paragraph{Global Speculation.} 

With global speculation enabled, messages destined to remote partitions are also sent immediately. 
Global speculation essentially moves all commit log updates out of critical path. It is particularly beneficial for workflows involving many hops between partitions. However, it requires a more involved protocol to ensure aborts are propagated correctly.

The sending partition keeps a record of the completed work items and the messages they have sent. When a work item is persisted, for each message sent before, it sends a confirmation message. The receiving partition knows that a message it receives is speculative until it receives that confirmation message; and the partition avoids persisting any work items that depend on such a speculative message until a confirmation is received.

But how are crashes handled? Note that when a partition crashes and recovers, it may no longer remember the work items it completed before the crash, so it cannot simply send abort messages for individual work items. Our current solution thus relies on using the commit log positions of partitions. Each speculative message is tagged with the commit log position of the work item that produces it. When a partition crashes and recovers, it broadcasts a recovery message to all partitions, which contains the recovered commit log position. When a partition receives a recovery message, it then "rewinds" its own commit log, by recovering from the closest preceding checkpoint, to a position that does not causally depend on aborted work items. It then broadcasts recovery messages of its own, to propagate aborts recursively. 
\section{Evaluation}
\label{sec:evaluation}

The goal of our evaluation is to study several aspects of DF and \sys. We start by describing the workflow applications (\S\ref{ssec:eval-benchmarks}). We then formulate the research questions (\S\ref{ssec:rq}), and present the results (\S\ref{ssec:eval-programmability}--\ref{ssec:eval-scaleout}).

\subsection{Workflows}
\label{ssec:eval-benchmarks}

\hide{
\kk{It might be a good idea to have a small illustration of each workflow in a figure to better understand their structure.}
\dajusto{+1, it's currently kinda abstract. Also, wouldn't this be better introduced \textit{before} the research questions and their results summary?}
\kk{I think that the benefit of having the questions and the result summary in the beginning makes it easy for someone to follow and know what to expect. Also being early helps someone get an idea of what we show without even looking at the evaluation in detail. IMO the specifics of the workflows are not relevant to that discussion could therefore be delayed.}
\dajusto{@kk I could put something together here if you could share the code that orchestrates these example.} \seb{We need to keep it short. Let's try to make the textual description as clear as possible. \dajusto{These are much clearer now, I'd say we're good on this.}}
}
We use five representative workflows that vary in complexity and execution characteristics. The first two workflows correspond to sequences of tasks, the third is a workflow that performs a transaction between two bank accounts and thus requires atomicity guarantees, and the other two workflows are taken from real applications, an image processing application, and a database snapshot obfuscation.

\resultpar{Hello Sequence.}
A very simple "hello world" workflow that calls three functions in sequence. Each function returns a hello message, and the workflow then returns the concatenation of those messages.

\resultpar{Task Sequence.}
A sequential workflow that initializes an object and then passes it through a sequence of processing steps. It is similar to the hello sequence, but the length of the sequence is not fixed, but given an an input parameter.
\hide{
\kk{Can we add something here that justifies why we picked the two sequence workflows? I think the reason is that they only have light computation in their tasks, therefore allowing us to measure the "true" performance of our engine and the existing solutions without noise from the execution of tasks.}
\dajusto{That explanation makes sense to me. You could add this bit in the first paragraph of this section. Something like: "The first two workflows are \textit{simple} sequential task compositions, they allow us to reason about the performance of Netherine by minimizing the noise from task execution time. The third is a ..." }
}


\resultpar{Bank Application.}
The workflow from Fig.~\ref{fig:transfer} that implements a reliable transfer of currency between accounts. This workflow showcases the capabilities of the Durable Functions programming model since it cannot be implemented with existing solutions.
%

\resultpar{Image Recognition.}
A workflow that recognizes objects in a given picture and creates a thumbnail for it. It is part of a bigger image processing application\footnote{Source at: \url{https://github.com/aws-samples/lambda-refarch-imagerecognition}}. 
The workflow performs the following steps, each of which is implemented as a separate AWS lambda. It first reads the image metadata from the S3 bucket where it is stored. If the image extension is supported, it filters out the unnecessary metadata and then runs two steps in parallel: one that performs object detection using Amazon Rekognition, and one that generates a thumbnail of the image. When the processes complete, it persists the filtered metadata in a DynamoDB table. The workflow repeatedly retries all steps until they succeed. 

\resultpar{Database Snapshot Obfuscation.}
This workflow is taken from a real application used for database snapshot obfuscation\footnote{Source at: \url{https://github.com/FINRAOS/maskopy}}. The workflow state machine contains 27 states that interact with a variety of AWS services. Some of the tasks that it calls include user authorization, creation of database snapshots, validation of the snapshots, obfuscation of the snapshots, and publishing the snapshots in a production environment. 


\begin{figure*}[t]
    \centering
    \includegraphics[width=\textwidth]{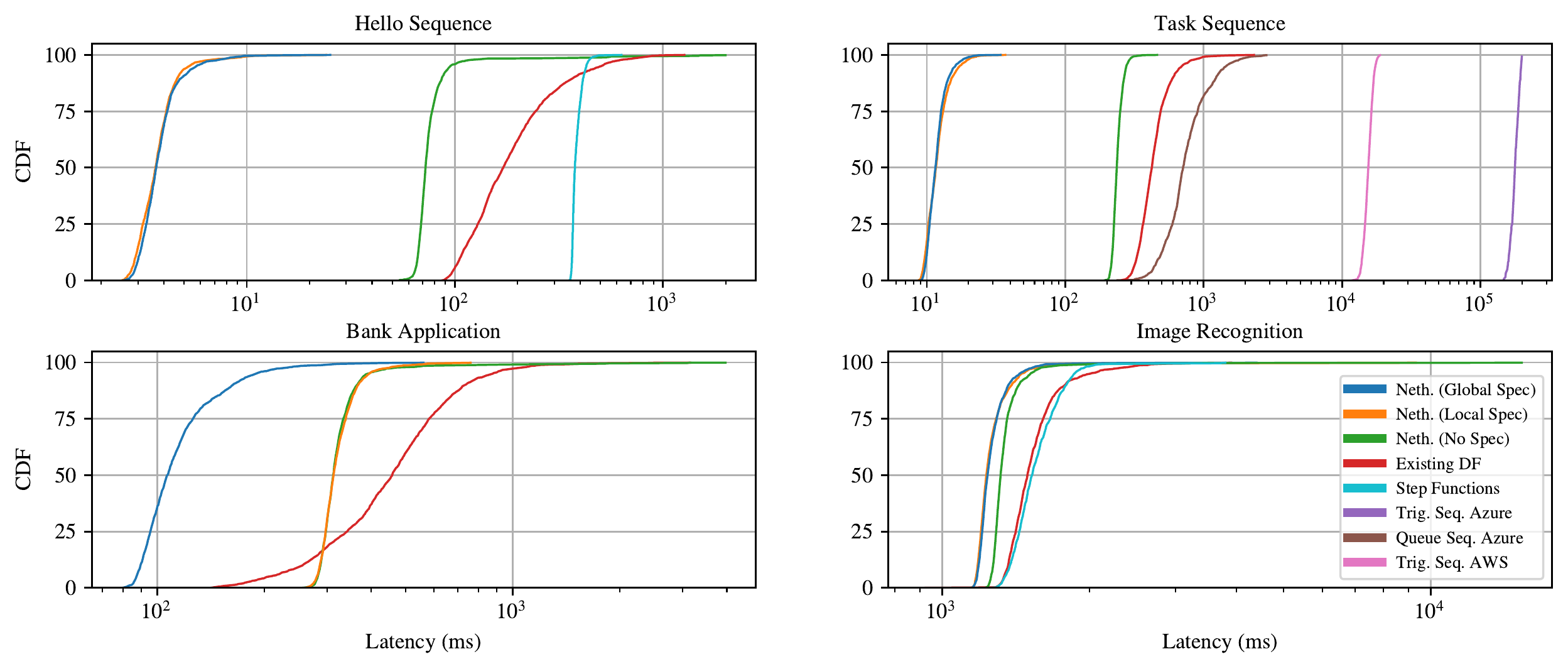}
    \caption{Latency measurements.}
    \label{fig:latencies}
\end{figure*}

\subsection{Research Questions}\label{ssec:rq}

We organize the evaluation and results according to the following questions:
\hide{
\kk{I personally prefer having the research questions before the workflows to help readers understand what they are about to read. Knowing the evaluation questions in advance helps the reader understand why these workflows are good choices for our evaluation.}
\seb{I like to separate the exposition of the benchmark from the results a bit more, by breaking it up with the research questions in between. It is not clear to me that the benchmarks should be chosen with respect to the research questions, they should rather appear like a random sample. }
}

\begin{description}[labelindent=0cm]
    \item[Q1] Does the DF programming model facilitate application development and maintenance?\str
    \item[Q2] How does \sys compare with existing solutions with respect to latency, i.e. the completion of a workflow?
    \item[Q3] How does \sys compare with existing solutions with respect to throughput, i.e. the number of workflows that it can execute in a period of time?
    \item[Q4] How does speculation improve latency and how does it impact throughput? 
    \item[Q5] Does \sys scale with the addition of available nodes in cases of high-load?
\end{description}

\paragraph{System infrastructure.} In all experiments other than the ones targeting AWS Step Functions, the system under test was run on a pool of Linux VMs on Azure Kubernetes Service, of type \verb+Standard_DS2_v2+ \cite{azure-vm-sizes}. The number of nodes was 4 (8 for the scale out experiment). Each node had 2 vCPU and a memory limit of 5GB. The queueing service was Azure EventHubs, which is roughly equivalent to Apache Kafka \cite{kafka}, with 32 partitions. The cloud storage was Azure storage GPv2, using premium tier for the FASTER Log Devices. The load was generated by a separate deployment of 20 load generator machines.

\subsection{Programmability Results (Q1)}
\label{ssec:eval-programmability}

To evaluate and compare the development experience when using DF, unstructured composition, or Step Functions, we tried to implement all the workflows from \S\ref{ssec:eval-benchmarks}.
\begin{figure}[t]
\begin{minted}[fontsize=\footnotesize]{csharp}
// Execute Image Type Check
string format = extractedMetadataJson.format;
if(!(format == "JPEG" || format == "PNG")) {
    throw new Exception($"image type {format}not supported");
}
\end{minted}
\caption{DF code that checks whether the input image is in a supported format.}
\label{fig:df-image-rec-if}
\end{figure}
%
\begin{figure}[t]
\begin{minted}[fontsize=\footnotesize]{csharp}
try {
    ...
} catch {
    // Catch errors by calling ErrorHandling
    string ErrorHandlingAndCleanupInput = 
            JsonConvert.SerializeObject(inputJson);
    await MakeHttpRequestSync(inputJson.ErrorHandlingURI, 
                              ErrorHandlingInput, context);
    return "Orchestration Failed!";
}
\end{minted}
\caption{DF code that does error handling for the snapshot obfuscation application.}
\label{fig:df-maskopy-error-handling}
\end{figure}

\resultpar{Task Sequence.}
With DF, the task sequence can be implemented using a straightforward for-loop that iteratively updates the target object by invoking the task with it. With unstructured composition, the sequence is also relatively simple, but requires that the user also manages and configures a storage or queue service. To our surprise, with Step Functions, it is not possible to express this workflow: the JSON schema for state machines does not support folds, i.e. loops with iteration dependencies. Encoding a loop by restarting the state machine does not work since the invocation API would return after the first iteration terminates. 


\resultpar{Image Recognition.}
In order to be as faithful as possible to the original Step Functions implementation, we implemented this workflow in DF by invoking the original Lambdas through their HTTP interface, only porting the workflow logic. The code in DF is 70 lines of standard C\#, while the state machine definition in Step Functions is 150 lines of JSON. An interesting difference is the implementation of a check whether the format of an image is supported. In Step Functions, this requires 24 lines of JSON \str 
compared to a 5 line if statement in DF (Fig.~\ref{fig:df-image-rec-if}).

\resultpar{Database Snapshot Obfuscation.}
The workflow in this application is by far the most complex. The state machine definition in Step Functions contains 27 states and is written using 700 lines of JSON; the DF version is more concise and easier to read, with 200 lines of C\# code. An important observation is that there is a lot of copied code in the Step Functions definition since it doesn't support function abstraction. Specifically, the error handling logic, written as 9 lines of JSON, is copied 12 times in the definition, while in DF we just wrap the orchestration with a single \texttt{try-catch}(Fig.~\ref{fig:df-maskopy-error-handling}).

\resultpar{Bank Application.}
The bank application simulates bank accounts and reliable money transfers between them. In DF, this is straightforward to implement using entities (Fig.~\ref{fig:entity}) and critical sections (Fig.~\ref{fig:transfer}). We have not yet figured out a satisfactory way of implementing this workflow using unstructured composition or Step Functions, as they do not provide the synchronization primitives needed for concurrency control.

\resultpar{Take away:}
DF supports a wider range of applications than Step Functions and unstructured composition, due to its support for workflows with rich control structure, entities, and critical sections. Defining workflows implicitly using a high-level language has several benefits compared to declarative definitions using JSON since it allows for features such as function abstraction and error handling. Maintenance is also improved since there is less copied code and less lines of code in general. Finally, using a high-level language the user gets to enjoy all of its benefits (libraries, type system, IDE support).

\subsection{Latency Results (Q2, Q4)}
\label{ssec:eval-latency}

In this section we conduct experiments using the workflows described in \Cref{ssec:eval-benchmarks} to evaluate \sys's latency and how it is improved by speculation.

\resultpar{Methodology.}
For all workflows except the snapshot obfuscation, requests are issued at a fixed, low rate (4--25 requests per second) for 3--5 minutes. We then compute the empirical cumulative distribution function (eCDF) of the  system-internal orchestration latency, i.e. the time it takes for an orchestration to complete, using the timestamps reported by the system. We chose to use the system-reported latency of workflows, as opposed to the client-observed latency, because not all clients provide a way to wait for the completion of a workflow.
\hide{
}

Latency results for four of the five workflows are shown in Fig.~\ref{fig:latencies}. For the snapshot obfuscation workflow, there is no appreciable performance difference between the implementations; the total latency (20-25 minutes) is dominated by executing the time-consuming tasks (taking a snapshot, obfuscating it, restoring the database from a snapshot, etc).

\resultpar{Unstructured Composition.}
Unstructured composition (using triggers and queues) can only be used to implement the Task Sequence workflow. As can be seen in Fig.~\ref{fig:latencies}, triggers\footnote{Blob in Azure and S3 in AWS.} suffer significantly higher latencies (\hl{x1000-x10000}) than \sys. Using queues for constructing sequential workflows performs better than triggers but \sys still achieves an order of magnitude lower latencies (\hl{median x61, 95th x91}).

\resultpar{Step Functions.}
Step Functions does not support the bank application and the task sequence so they are not included in that experiment. For the other two workflows \sys achieves better latencies (hello sequence: \hl{median x104, 95th x75}). An important take-away is that \sys achieves lower latency in the image recognition experiment even though \sys is deployed on Azure and invokes AWS lambdas as its tasks using their HTTP interfaces, while AWS Step Functions invoke the lambdas directly (avoiding both the network back and forth and the HTTP overhead).
\hide{\kk{Add a note that Step Functions code is not available and thus we cannot explain the results.}}

\resultpar{Durable Functions.}
Compared to the existing implementation of Durable Functions, \sys achieves better latency in all experiments, even without speculation. The optimized \sys implementation achieves \hl{x38}, \hl{x4.3}, \hl{$17\%$}, improvements in median and \hl{x43}, \hl{x4.7}, \hl{$29\%$} improvements in 95th percentile latency than the existing implementation in the task sequence, bank, and image recognition workflows respectively.

\resultpar{Speculation Benefits (Q4).}
The benefits of speculation are apparent in all plots of Fig.~\ref{fig:latencies}. In general, the improvement is cumulative, with two exceptions: local speculation does not improve latency for the Bank Application since there is a lot of communication among workflows and entities, and global speculation does not improve latency for task sequence and image recognition since their workflows stay within partitions. In image recognition, the speculation benefits are small because the biggest factor of the workflow latency is the execution time of the image recognition. In total, median latency for the sequence experiment is improved  by \hl{x21} (95th \hl{x17}) with speculation, the median latency for the image recognition experiment is improved by \hl{$6\%$} (95th \hl{$5\%$}) due to speculation, and finally the median latency for the bank experiment is improved by \hl{x3} (95th \hl{x2}) using global speculation.

\resultpar{Take away:}
\sys achieves better latencies than all other solutions in all of our experiments. Speculation significantly improves \sys's latency. For a workflow taken from an AWS application, \sys achieves better latency than Step Functions even though it pays communication and HTTP costs due to being deployed in Azure and calling stateless functions deployed in AWS.



\subsection{Throughput Results (Q3, Q4)}
\label{ssec:eval-throughput}


\begin{figure}[t]
    \centering
    \includegraphics[width=\columnwidth]{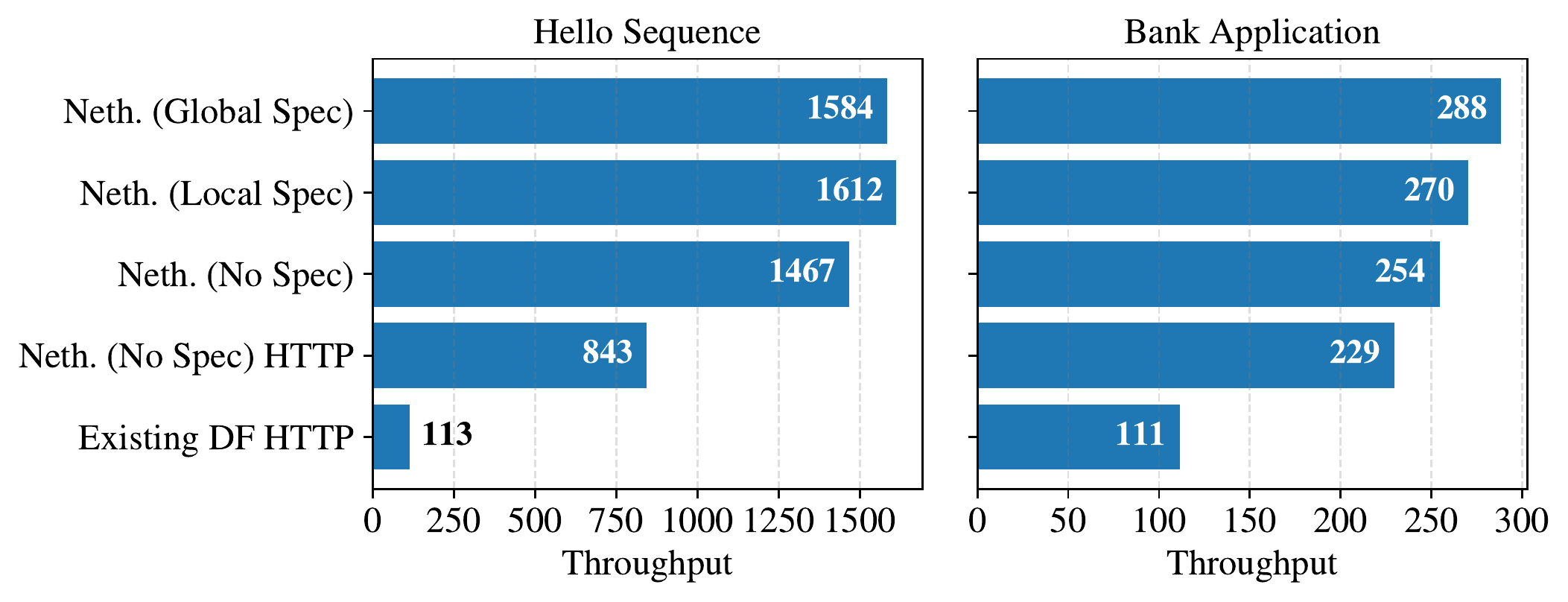}
    \caption{Throughput Measurements. \str
    }
    \label{fig:throughputs}
\end{figure}

In this section we conduct experiments to evaluate \sys's throughput and how is it impacted by speculation.

\resultpar{Methodology.}
For the throughput experiments, we control the load by controlling the number of request loops that are running on the load generators. We determine a suitable load level by ramping up the load until we can visually discern saturation, indicating that a further load increase will not improve throughput. We then keep that load steady for a minute and compute the average throughput.

We only compare against the existing DF implementation because it is available on Github\footnote{\url{https://github.com/Azure/azure-functions-durable-extension}} and thus we could deploy it with the exact same resources as \sys.

We did not include image recognition and snapshot obfuscation since their throughput limits are bounded by the throughput limits of external services that they use\footnote{AWS Rekognition has a limit of 50 invocations per second and the snapshot obfuscation workflow takes 20-25 minutes to complete.}. We did not include throughput measurements for task sequence because its results are very similar to the Hello Sequence. Throughput results are shown in Fig.~\ref{fig:throughputs}.

\resultpar{Durable Functions.}
The HTTP plots correspond to executions where the invocations where done through HTTP, consuming some resources.
\sys without speculation improves the throughput over the existing DF implementation by x7.5 for hello sequence and by x2 for the bank application.
Throughput improvement for the bank application is smaller, presumably because there is much inter-partition traffic and less batching per node.

\resultpar{Speculation (Q4).}
To measure speculation improvement on throughput more accurately, we invoke the workflows without HTTP. Speculation slightly improves throughput in both experiments: for Hello Sequence ($10\%$ with local, $8\%$ with global), for Bank Application ($6\%$ with local, $13\%$ with global). It is not immediately clear why global speculation improves throughput of the bank application, as it performs strictly more work per orchestration. We believe the reason is that the much lower latency (almost \hl{x5}) means each workflow spends less time in the system, leading to emptier queues, less memory consumption, and less GC overhead.

\resultpar{Take away:}
\sys achieves close to \hl{x8} the throughput of the existing DF implementation. Speculation does not negatively impact throughput, but slightly improves it.

\subsection{Scale-out Results (Q5)}
\label{ssec:eval-scaleout}

\begin{figure}
    \centering
    \includegraphics[width=\columnwidth]{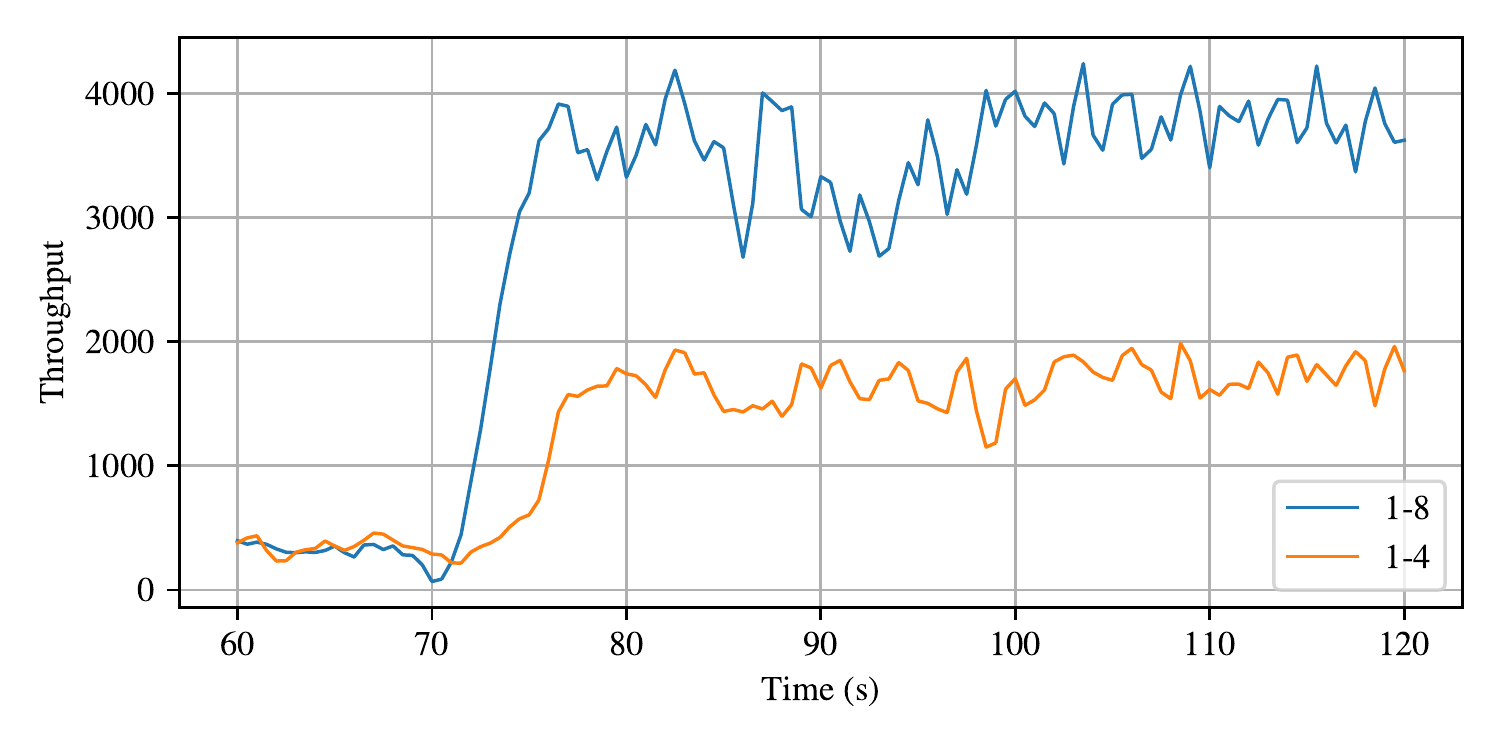}
    \caption{Scale out throughput plot. Each line is averaged over 3 runs to reduce noise.}
    \label{fig:scaleout}
\end{figure}

In this section we conduct an experiment to evaluate if \sys can scale out with the addition of nodes.

\resultpar{Methodology.}
For this experiment, as before in \Cref{ssec:eval-throughput}, the load generators emit a fixed load that can saturate the throughput of the full configuration (4 or 8 compute nodes). We start the experiment with all 32 partitions located on a single node, while the other nodes are unused. After running for 70 seconds, we re-balance the partitions across all available nodes. Specifically, for the 4-node experiments, we move 24 of the 32 partitions, and for the 8-node experiments, we move 28 of the 32 partitions. To reduce noise we repeated this 3 times for each series and computed the average.

\resultpar{Observations.} The results for the hello sequence workflow without speculation are shown in \Cref{fig:scaleout}. They show that \sys can scale out with the addition of compute instances. It reaches peak throughput after around 7 seconds (after the scale-out decision at 70s) in both cases. 
\str

\section{Related Work}


%
\paragraph{Workflows.} 
Many systems have acknowledged the challenges of providing reliable execution guarantees for long-running workflows. Most follow the declarative approach: Netflix's Conductor~\cite{conductor}, Zeebe~\cite{zeebe}, and AWS Step Functions~\cite{aws-step-functions} use a JSON schema for authoring workflows, and Fission Workflows~\cite{fission} use YAML.  Apache Airflow~\cite{airflow} and its productization, Google Cloud Composer~\cite{cloudcomposer}, and Fn Flow~\cite{flow}, are somewhat more code-based, as the schema is constructed in code. However, unlike DF, they do not simply adopt the standard control flow semantics of the host language. Also, none of these systems offer state and synchronization features comparable to entities or critical sections.

\paragraph{Actors.}  Entities in DF, and the instances in the computation model, are inspired by traditional actor systems like Erlang~\cite{armstrong1997development} or Akka~\cite{haller2012integration}, and especially the virtual actors of Orleans~\cite{bykov2011orleans}. The latter support persistence, but may lose actor messages, guaranteeing only "at-most-once" delivery. Similarly, the execution guarantees for Cloudflare's Durable Objects\footnote{Durable Objects closed beta was announced on September 28th, 2020.}~\cite{introduction-to-durable-objects, using-durable-objects} and Lightbend's Akka Serverless\footnote{Akka Serverless was formerly known as Lightbend CloudState.}~\cite{cloudstate}, apply only to a single object; they do not provide causal consistency guarantees or synchronization primitives that span multiple objects, like DF orchestrations.

\paragraph{Engine.} The \sys architecture is inspired by Ambrosia's virtual resiliency \cite{ambrosia}, with the partitions corresponding to immortals. However, instead of a single queue, \sys uses a separate commit log and input queue, which reside in different external services. Also, \sys keeps cold state in storage by virtue of Faster \cite{chandramouli2018faster}, and supports speculation. 

\paragraph{Serverless Computation Model.} The need to augment FaaS with support for state and synchronization has been acknowledged by both the research and industrial communities~\cite{DBLP:conf/cidr/HellersteinFGSS19,DBLP:journals/corr/abs-1902-03383,DBLP:conf/cidr/Schleier-Smith19,8481652}. Jangda et al. \cite{jangda-et-al} present a formal model for FaaS and explain its limitations. They also show how to compose functions using a language called \verb+SPL+. However, unlike our work, they do not combine state, messages, and functions into a single serverless model with reliable and causally consistent execution guarantees. 

Cloudburst~\cite{sreekanti2020cloudburst}, on the other hand, has a similarly unified computation model, and can also, like Netherite, execute efficiently and guarantee causal consistency. However, its programming model does not allow for arbitrary dynamic task-parallel code, like DF, but supports execution of statically specified DAGs only. Similarly, recent work \cite{fault-tolerance-shim,transactional-causal-consistency} investigates how to guarantee causal consistency for serverless applications, but for a workload of transactions over replicated data, not message-passing workflows. The difference is that in our model, only each message-processing step, not the entire workflow, executes transactionally. 

%
%
Kappa~\cite{zhang2020kappa} proposes a programming framework for serverless that addresses two issues with serverless functions: the execution time limit and the lack of coordination between lambdas. Kappa is very similar to DF in that it also offers a high-level language programming environment. The main difference with DF is that it doesn't support some advanced features such as error handling and critical sections.
%
PyWren~\cite{jonas2017occupy}, mu~\cite{fouladi2017encoding}, and gg~\cite{fouladi2019laptop} all propose simple programming frameworks for developing parallel serverless applications by exploiting the scalability of serverless functions. In contrast, DF is a complete programming solution that supports advanced features (arbitrary composition, critical sections) and also offers strong reliability guarantees.
\section{Conclusion and Future Work}

We have devised, explained, implemented, and evaluated a comprehensive platform for authoring and executing serverless workflows. We have shown that, compared to prevailing approaches, (1) the DF programming model improves the developer experience and broadens the scope of supported applications, and (2) the Netherite architecture improves the performance across the board, by orders of magnitude in some cases. Our work enables the development of full-featured, stateful, serverless applications that extend far beyond the scope of the original FaaS concept.

In future work, we would like to explore how to extend CCC to external services, and how to further improve Netherite by smarter scheduling of tasks and steps. \hide{\dajusto{grammar nit: replace "by smarter" with "via smarter" or "with smarter"}}

\bibliographystyle{plain}
\bibliography{bib}

\end{document}